\documentclass[prl,reprint]{revtex4-1}
\usepackage{graphicx}
\usepackage{bm,braket,color,ulem}
\usepackage{amssymb}
\usepackage{txfonts}

\begin{document}

\title{Superconducting fluctuation in FeSe investigated by precise torque magnetometry}

\author{Hideyuki Takahashi$^{1,2}$, Fuyuki Nabeshima$^3$, Ryo Ogawa$^3$, Eiji Ohmichi$^4$, Hitoshi Ohta$^5$, Atsutaka Maeda$^3$}

\affiliation{
$^{1}$Organization for Advanced and Integrated Research, Kobe University, 1-1 Rokkodai-cho, Nada, Kobe 657-8501, Japan\\
$^{2}$Japan Science and Technology Agency, PRESTO, 4-1-8 Honcho, Kawaguchi, Saitama, 332-0012, Japan\\ 
$^{3}$Department of Basic Science, the University of Tokyo, 3-8-1 Komaba, Meguro-ku, Tokyo 153-8902, Japan\\
$^{4}$Graduate School of Science, Kobe University, 1-1 Rokkodai-cho, Nada, Kobe 657-8501, Japan\\
$^{5}$Molecular Photoscience Research Center, Kobe University, 1-1 Rokkodai-cho, Nada, Kobe 657-8501, Japan\\
} 

\date{\today}

\begin{abstract}
We investigated the superconducting fluctuation in FeSe, which is assumed to be located in the BCS--BEC crossover region, via magnetic torque measurements. 
In our method, the absolute cantilever displacement is measured by detecting the interference intensity of the Fabry--Perot cavity formed between the cantilever and optical fiber.
Our findings are totally different from the results of the previous torque magnetometry using a piezoresistive cantilever; the "giant" fluctuation diamagnetism related to the BCS--BEC crossover does not exist.
Instead, a considerably smaller fluctuation signal originating from the vortex liquid was observed that showed a qualitatively similar behavior to those in cuprate superconductors.
We also discuss the inconsistency between our torque data and the existence of a pseudogap proposed by an NMR experiment.
\end{abstract}

\maketitle

In conventional superconductors, the superconducting fluctuation (SCF) is observed within in a very limited temperature range above $T_c$. 
However, in high $T_c$ cuprate superconductors, owing to the short coherence length and low dimensionality, fluctuation effect is accessible through various experimental techniques, eg. Nernst effect~\cite{Xu2000}, magnetic torque~\cite{Wang2005, Li2010}, and microwave to terahertz spectroscopy~\cite{Ohashi2009, Nakamura2010, Bilbro2011, Nakamura2012}.
As SCF is the preceding phenomena of superconductivity, it is important to investigate in detail the characteristics of SCF to clarify the mechanism of unconventional superconductivity.

After an angle-resolved photoemission spectroscopy revealed a very large superconducting gap to Fermi energy ratio $\Delta/E_F\sim 0.5$ in FeSe$_x$Te$_{1-x}$, iron chalcogenide superconductors have been attracting a great deal of interest~\cite{Lubashevsky2012, Okazaki2014}. 
From the analogy of ultracold Fermi gas, they are said to be located in Bardeen-Cooper-Schrieffer(BCS) to Bose-Einstein-Condensate(BEC)  crossover region~\cite{Randeria2010}. 
It would be very interesting if these two very different physical systems can be described via a common theoretical framework.
To explore the anomalous superconducting property owing to BCS-BEC crossover, measuring the SCF effect has been proposed.
As the temperature decreases, the system enters the pseudogap (PG) region in which strongly interacting electrons make so-called preformed-pairs prior to the superconducting transition. 
It is naive to expect that SCF is enhanced in the PG phase.

SCF in iron chalcogenides has been most intensively studied on stoichiometric FeSe of which high quality single crystal can be synthesized~\cite{Kasahara2014}.
However, a consensus has not yet been reached. 
It has been reported that the giant diamagnetic signal was observed below 20K by cantilever magnetometry~\cite{Kasahara2016}. 
An NMR experiment showed a weak anomaly in the temperature dependence of the spin-lattice relaxation rate $1/T_1$~\cite{Shi2018}. 
On the other hand, such a large fluctuation signal was not observed in the later magnetization measurement~\cite{Yuan2017}. 
Likewise, according to the microwave spectroscopy on FeSe$_{0.5}$Te$_{0.5}$, SCF has not been observed above 1.1$T_c$ in a zero magnetic field limit~\cite{Nabeshima2018}.
This discrepancy suggests the difficulty experienced in the precise measurement of the SCF signal.
Therefore, an improved experimental technique is crucially needed to overcome the current situation.

\begin{figure}[tb]
	\begin{center}
		\includegraphics[width=0.9\hsize]{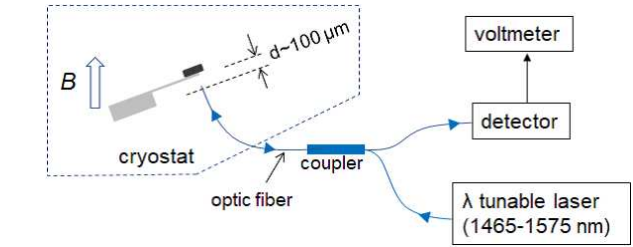}
	\end{center}
\caption{Schematic of the experimental setup.}
\label{experimental}
\end{figure}

\begin{figure*}[tb]
	\begin{center}
		\includegraphics[width=0.85\hsize]{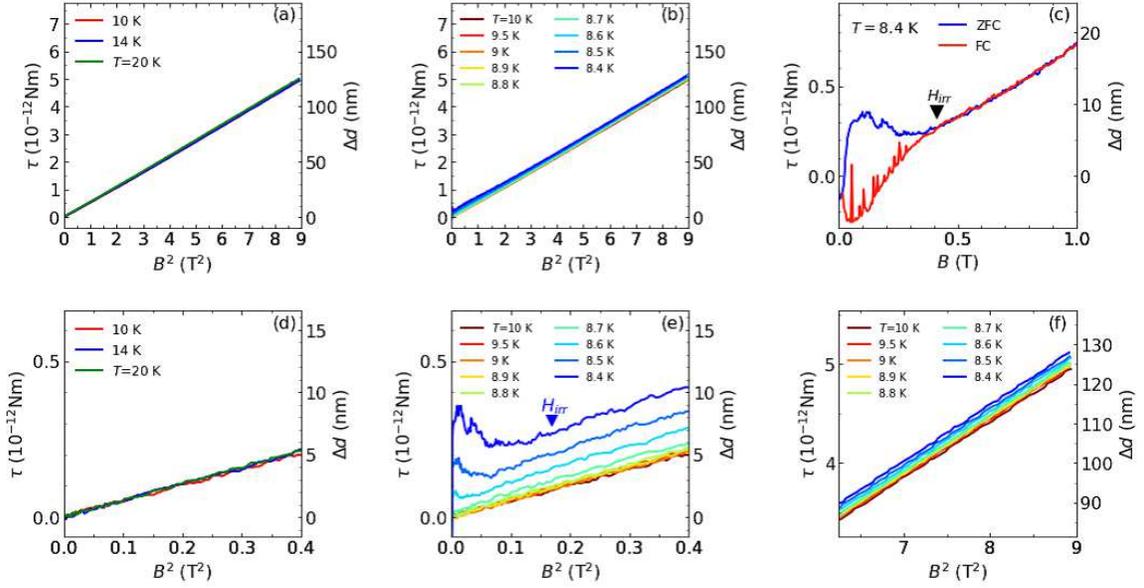}
	\end{center}
\caption{Cantilever displacement in the magnetic field at (a) $T=$10-20 K and (b) 8.4-10 K in ZFC condition. (c) Hysteresis between ZFC and FC data at $T$=8.4 K. The closed triangle shows the irreversibility field $H_{irr}$. The data below 1 T in (a) and (b) are respectively replotted in (d) and (e). The data between 2.5 T and 3 T in (b) are expanded in (f).}\
\label{fig2}
\end{figure*}

In the previous torque magnetometry study, a commercial piezoresistive cantilever (PRC400, HItachi Hi-tech corp.) was used~\cite{Kasahara2016}.
This type of cantilever detects its displacement as a change in the piezoresistance $\Delta R_p$ of its resistive path (P$_1$ with total resistivity $R$) implanted in the cantilever legs.
As it can easily convert magnetic torque into electric signals, it is a very useful tool particularly in low-temperature experiments~\cite{Ohmichi2002}.
However, we would like to point out some possible issues in using piezoresistive cantilever for the quantitative measurement of small signals such as SCF.
$R$ depends on both temperature $T$ and magnetic field $B$.
In a typical case, $R(T, B)$ changes 1-2\% in the ranges of $5\ \mathrm{K}<T<30\ \mathrm{K}$ and $0\ \mathrm{T}<B<10\ \mathrm{T}$.
This change is in most cases larger than $\Delta R_p$ caused by the magnetic torque.
Although the Wheatstone bridge circuit including the compensation resistance (P$_2$ having almost the same property as P$_1$) is usually used to extract $\Delta R_p$, this method cannot completely exclude the influence of $T$-/$B$-dependent resistivity.
Therefore, it is quite difficult to quantitatively evaluate the extremely small SCF signal, while it is useful in qualitative measurements such as de Haas-van Alphen effect~\cite{Ohmichi2002} and magnetic resonance measurement~\cite{HT2015}.

In this letter, we present the data of SC-induced magnetic torque using a more precise method.
The experimental setup is shown in Fig.1.
We measure the cantilever displacement $\Delta d$ induced by the magnetic torque, $\bm{\tau}=V\bm{M}\times\bm{B}$ ($V$ is the sample volume and $\bm{M}$ is the magnetization per unit volume) by forming a low-finesse Fabry--Perot interferometer between the cantilever and cleaved end of the optic fiber. 
A fiber-coupled laser (81989A, Agilent) was used as the laser source~\cite{Smith2009}.
The reflected beam from the interferometer was divided by the photocoupler (10201A-90, Thorlabs), and fed into a photodetector (81636B, Agilent).
The interference signal, $I$, can be expressed as a function of the cantilever-fiber distance $d$ and the laser wavelength $\lambda$,
\begin{equation}\label{eq1}
I=\frac{I_{MAX}+I_{MIN}}{2}-\frac{I_{MAX}-I_{MIN}}{2}\cos (\frac{4\pi d}{\lambda}),
\end{equation}
where $I_{MAX}$ ($I_{MIN}$) is the maximum (minimum) interference intensity. 
As the wavelength of the 81989A laser source can be varied in the range of 1465--1575~nm, the interferometer can be tuned to the optimal point where $d/\lambda =1/8+n/4\ (n=0,\ 1,\ 2\cdots)$ without using a mechanical positioner.
A great advantage of this method is that it uses $\lambda$ as a reference scale.
Therefore, the absolute displacement can be derived without complicated calibrations.

We used FeSe single crystals grown by the chemical vapor transport method~\cite{Bohmer2013}.
The residual resistivity ratio was determined to be over 50, which is comparable to the one used in the previous torque study~\cite{Kasahara2016}.
The platelet crystal having dimensions of $80\times 140\times 30\ \mathrm{\mu m^3}$ was attached on an AFM-cantilever (PPP-CONTSCR, Nanosensors$^{\mathrm{TM}}$), having a spring constant of $k=0.2$ N/m and dimensions of $l\times w\times t=225\times48\times 1\ \mathrm{\mu m^3}$, using a small amount of epoxy glue.
The sample temperature was readout by the Cernox sensor placed right next to the sample.
To avoid temperature difference between the sample and sensor, we performed experiments in a $^4$He gas atmosphere.
The peak-to-peak vibration noise of the cantilever was reduced to about 0.1 nm by minimizing the $^4$He gas flow rate.

The magnetic torque is caused by the anisotropy of the magnetic susceptibility $\Delta \chi=\chi_c-\chi_{ab}$, where $\chi_c (\chi_{ab})$ is the magnetic susceptibility along $c (ab)$-axis.
When the sample $c$-axis is tilted from the $B$ direction, it is expressed as $\tau=kl\Delta d=\frac{1}{2\mu_0}\Delta \chi VB^2\sin 2\theta$.
When $\Delta d< \lambda/4$, $\Delta d$ can be directly derived from Eq.~(\ref{eq1}). 
On the other hand, when $\Delta d> \lambda/4$, it should be considered that $I_{MAX}$ ($I_{MIN}$) is no longer constant.
The large cantilever deflection changes the angle between the incident laser beam and cantilever surface, then it decreases the interference amplitude $\frac{I_{MAX}-I_{MIN}}{2}$.
In such cases, $\Delta d$ is obtained by counting the maximum and minimum of interference intensity which appear for every $\lambda/4$ displacement~\cite{HT2016}.

\begin{figure}[tb]
	\begin{center}
		\includegraphics[width=1\hsize]{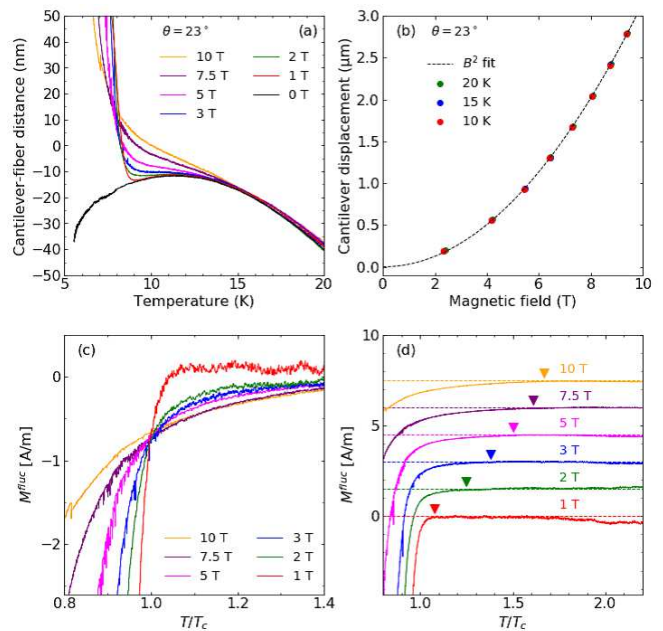}
	\end{center}
\caption{(a) $d$ vs $T$ in the magnetic field for $\theta=23^{\circ}$. The zero-field data shows the background signal due to the thermal expansion of the interferometer. (b) $B$-dependence of the cantilever displacement. This data was taken by counting the interference intensity maximum/minimum which appears for every $\lambda/4$ displacement. The dashed line shows the $B^2$ fit.  (c) Fluctuation diamagnetization derived from the data in (a). (d) Same data as (c), except the data are shifted for clarity. The closed triangles show the SCF onset temperature.}
\label{fig3}
\end{figure}

Figure~\ref{fig2}(a) and (d) shows the magnetic field dependence of the cantilever deflection between 10 K and 20 K. 
The $c$-axis was tilted from $B$ direction by $\theta=10^{\circ}$.
We confirmed that the direction of the displacement exhibits $\chi_c > \chi_{ab}$ as reported in previous studies.
No significant temperature dependence was observed between 10 and 20 K.
The data are well fitted by $B^2$.
As shown later in Fig.~\ref{fig3}(b), $\tau$ is almost perfectly proportional to $B^2$ throughout the measured magnetic field range up to 10 T.
This suggests that the magnetization is linear in $B$.
At the same time, this assures the excellent linearity of the cantilever's spring constant at least in the displacement range of up to $3\ \mathrm{\mu m}$. 

The magnetic torque exhibits an abrupt change after it enters the superconducting state as shown in Fig.~\ref{fig2}(b), (c), and (e).
In the low field region, we observed the hysteretic behavior due to the vortex pinning effect between the zero-field-cooled and field-cooled data.
We define $T_c=8.7\ \mathrm{K}$ as the temperature where the hysteresis becomes observable.
In general, there is a first-order melting transition of the vortex lattice above the irreversibility field $H_{\mathrm{irr}}$ showing a discontinuous change in the magnetization.
However, we did not find any anomaly in torque data.
This is probably because the melting transition is very close to $H_{\mathrm{irr}}$ in iron-based superconductors~\cite{Mak2013}. 
In this case, it is reasonable to consider that the smooth magnetic field dependence above $H_{\mathrm{irr}}$ results from the SCF in a vortex liquid (VL) state.
The data between 2.5 and 3 T in Fig.~\ref{fig2}(b) is expanded in Fig.~\ref{fig2}(f).
We can see the gradual increase of the torque signal manifesting such that the VL phase exists in a considerably higher field than $H_{irr}$ even near $T_c$.
Similar vortex-like signal has been observed in cuprate superconductors as well~\cite{Xu2000, Wang2005}.
This is attributed to the large phase fluctuation of the Cooper-pair wave function.
Extensive studies have been performed for various compounds and it has been discovered that the VL survives up to the temperature several times higher than $T_c$~\cite{Li2010}.

To determine how high in $T$ the diamagnetic signal extends above $T_c$ in FeSe, we extended the experiment to a higher field region and measured the $T$-dependence of $\tau$ with higher resolution.
In temperature-swept measurements, we should be careful as the thermal expansion of the measurement system changes $d$ and causes a background signal.
We confirmed that the background signal is sufficiently small and reproducible below 20 K from the data with $B=0$ (Fig.~\ref{fig3}(a)).
In the magnetic field, we observed the deviation from the background signal from above $T_c$.
The direction of the deviation is consistent with the VL signal in Fig~\ref{fig2}.
We extracted the effective fluctuation diamagnetization $M^{\mathrm{fluc}}$ by subtracting the background signal as $M_{\mathrm{fluc}}(T, B)=kl\{d(T, B)-d(T, 0)\}/VB\sin\theta$ and determined the onset temperature $T_{\tau}^\ast$ (Fig.~\ref{fig3}(d)).
Although there is an uncertainty of about $\pm 0.05T_c$, it is clear that $T_{\tau}^\ast$ increases as magnetic field increases and reaches 1.6$T_c$ at 10 T.
Below $T_c$, $M^{\mathrm{fluc}}$ at each magnetic field cross each other in the narrow temperature region (inset in Fig.~\ref{fig3}(c)).
Such behavior is widely observed in cuprate superconductors, which can be theoretically reproduced using two- or three-dimensional lowest-Landau-level approximation~\cite{Welp1991,Li1992,Tesanovic1992,Gao2006}.
In addition, Adachi et al. recently calculated the fluctuation diamagnetism in the BCS and BCS--BEC crossover region, and demonstrated that the magnetization crossing can occur in both regions~\cite{Adachi2017}.
These studies suggest that the crossing phenomenon is a universal property of SCF. 
Therefore, this observation provides strong evidence that our data concerning SCF diamagnetism is reliable. 

We emphasize that the signal we observed has totally different features, both qualitatively and quantitatively, from the one observed in the previous magnetic torque measurement by Kasahara et al. as follows;
(i) In Ref~\cite{Kasahara2016}, $\tau$ shows pronounced $T$-dependence. 
As $T$ approaches $T_c$, $\tau$ changes $\sim 10$ \% between $T_c$ and 30 K.
This signal was assumed to be the evidence of the giant SCF in BCS--BEC crossover region. 
However, our result unambiguously shows the absence of such a large fluctuation signal (Fig.~\ref{fig2}(a)).
(ii) In Ref~\cite{Kasahara2016}, the fluctuation signal is enhanced below 1 T, which causes the nonlinear $B$ dependence of the diamagnetic signal.
In contrast, our result shows that the SCF diamagnetism becomes pronounced on increasing $B$.
In addition, this SCF is not large enough to change the global $B$-dependence of sample magnetization.
As shown in Fig.~\ref{fig3}(a), even at 10 K and 10 T where the SCF signal is the largest, $M^{\mathrm{fluc}}$ contributes to the cantilever displacement of about 13 nm.
This value corresponds to only 0.4 \% of the total torque signal as shown in Fig.\ref{fig3}(b).
Such discrepancies are too large to be attributed to the difference in the crystal quality.

Furthermore, the reason why Yuan et al. could not observe the fluctuation by SQUID magnetometry can be explained by the weakness of the signal~\cite{Yuan2017}.
In their data, the longitudinal component of $M$, $M_z$, contains Curie-type signal from impurity. This makes it hard to separate the small SCF signal from the total magnetization.

\begin{figure}[tb]
	\begin{center}
		\includegraphics[width=0.8\hsize]{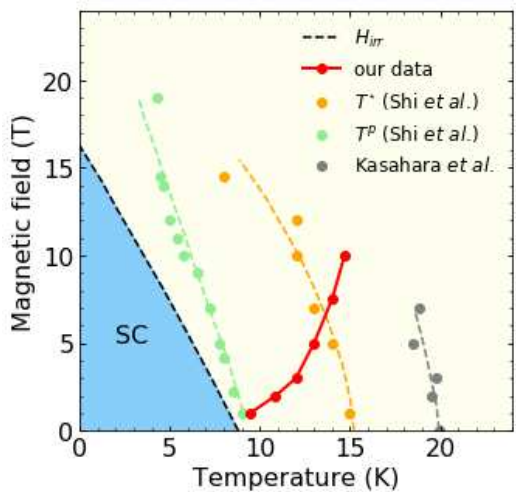}
	\end{center}
\caption{The onset temperature of SCF determined from $\tau$ is plotted on the $B$-$T$ phase diagram. Characteristic temperatures reported in previous studies are also plotted~\cite{Kasahara2016,Shi2018}. }
\label{fig4}
\end{figure}

In Fig.~\ref{fig4}, we summarize $T_\tau^{\ast}$ in the $B$-$T$ phase diagram together with the characteristic temperatures reported in previous studies.
$T_\tau^{\ast}/T_c$ in FeSe is considerably high and close to the values determined in cuprate superconductors.
In cuprate superconductors, the strong SCF is attributed to the high $T_c$ and two-dimensionality.
In this sense, FeSe is a unique superconductor exhibiting a similar fluctuation phenomena in spite of low $T_c$ and almost three-dimensional physical property.
Nevertheless, considering the large $\Delta/E_F$ value, SCF in FeSe is not anomalously large.
In the framework of Ginzburg-Landau theory, the temperature region where thermodynamic quantities require correction considering SCF is given by the Ginzburg-Levanyuk number $G_i$.
In the case of a clean three-dimensional superconductor, $G_i$ is approximated by $80(T_c/T_F)^4$~\cite{Larkin}.
This value is several orders smaller than unity in most cases.
However, for FeSe, $G_i$ is estimated to be 0.1-1 because of the large $\Delta/E_F$, which is consistent with our experimental results.

Finally, we discuss the inconsistency between our torque data and the existence of a PG phase proposed by the NMR experiment.
The emergence of PG phase accompanied with preformed pairs is a key feature of the BCS--BEC crossover superconductor.
Shi et al. have observed that $T_1^{-1}$ shows the magnetic field dependence from above $T_c$ ($T^{\ast}$ in fig.~\ref{Fig4})~\cite{Shi2018}.
Under the assumption that superconductivity disappears at $B=19\ \mathrm{T}$, they attributed the lower $T_1^{-1}$ at low magnetic fields to the decrease of the electronic density of states owing to the preformed pair formation.
However, we have shown that SCF becomes more pronounced under the magnetic field and SCF probably survives even at 19 T.
Therefore, their conclusion should be reconsidered.
In fact, we did not observe any correlation between $T_\tau^{\ast}$ and PG-like phase determined by Shi et al.
The torque data does not show any anomaly when crossing the boundary between the proposed PG phase and the normal metal state.

It should be noted that the SCF effect does not necessarily reduces $T_1^{-1}$.
Several mechanisms that increases $T_1^{-1}$ in the SCF region have been proposed to explain the similar behavior seen in optimally-doped cuprates, such as pairing symmetry-dependent fluctuations~\cite{Mitrovic1999} and the fluctuations enhanced by Landau-leval degeneracy~\cite{Zheng2000}. 
Irrespective of the reason, the interpretation that SCF in the magnetic field gives an additional relaxation rate is consistent with our results.
We expect that if the NMR data are reanalyzed according to this scenario, $T_1^{-1}$ will give a similar SCF onset temperature as $T_\tau^{\ast}$.

In view of the above discussions, SCF in FeSe is not as anomalous as proposed by previous torque and NMR studies.
The large value of $\Delta/E_F$ certainly enhances the SCF, which enables us to observe its effect up to 1.6$T_c$.
However, SCF in FeSe exhibits qualitatively similar behavior to that in cuprate and no conceptually new phenomena was observed.
Our result suggests that a large $\Delta/E_F$ alone is not sufficient to cause an equivalent phenomena as that of the ultracold Fermi gas.

This study was partly supported by the Grant-in-Aid for Young Scientists (18K13501) from the Japan Society for the Promotion of Science and the Kansai Research Foundation for technological promotion.

\end{document}